\documentclass[aps,prfluids, reprint,superscriptaddress, longbibliography]{revtex4-2}

\usepackage{graphicx}
\usepackage{amsmath,amssymb}
\usepackage{hyperref}
\hypersetup{colorlinks=true,allcolors=blue}
\newcommand{\bfu}{\mathbf{u}}
\newcommand{\bfw}{\boldsymbol{\omega}}
\newcommand{\curl}{\nabla\times}

\newcommand{\E}{E}
\newcommand{\Z}{Z}
\newcommand{\D}{D}
\newcommand{\PiC}{\Pi}

\newcommand\HL[1]{{\color{black}#1}}

\begin{document}

% ---------------- Title & authors -----------------
\title{Spectrum of the Curl of Vorticity as a Precursor to Dissipation in 3D Taylor--Green Turbulence}

% TODO: Replace with real authors and affiliations
\author{Satori Tsuzuki}\email{tsuzukisatori@g.ecc.u-tokyo.ac.jp}
\affiliation{Research Center for Advanced Science and Technology, The University of Tokyo}

\date{\today}

% ---------------- Abstract -----------------
\begin{abstract}
Predicting when a three-dimensional turbulent flow reaches its dissipation peak is essential for both theory and adaptive algorithms in simulations and experiments.
Using direct numerical simulations (DNSs) of the Taylor--Green vortex (TGV) at resolutions of $256^3$--$1024^3$, we introduce and test a small-scale weighted diagnostic: the spectrum of $|\nabla \times \boldsymbol{\omega}|^2$ (with $\boldsymbol{\omega}=\nabla \times \mathbf{u}$), which, for incompressible flow, is equivalent to a $k^4$-weighted energy spectrum.
We show that the peak wavenumber of this spectrum, $k_{{\rm peak}}[\,|\nabla \times \boldsymbol{\omega}|^2\,]$, advances rapidly to intermediate-small scales and then levels off \emph{before} the dissipation rate $\varepsilon(t)=\sum_k 2\nu k^2 E(k)$ reaches its maximum.
Across all resolutions, we observe robust temporal ordering $t_k<t_\varepsilon<t_\Pi$, where $t_k$ marks the onset of the rapid rise of $k_{{\rm peak}}[\,|\nabla \times \boldsymbol{\omega}|^2\,]$, $t_\varepsilon$ is the time of the maximal $\varepsilon(t)$, and $t_\Pi$ is when the cumulative flux $|\Pi(K)|$ attains its largest peak scale.
This early-warning signal correlates with the morphological transition to filament-dominated structures visible in $Q$-criterion isosurfaces and is consistent with integral-scale trends ($L_{{\rm int}},\lambda,\eta$).
The diagnostic is simple to compute from standard DNS data and highlights the incipient formation of high-curvature structures, where viscosity acts most strongly.
\end{abstract}

\maketitle

% ---------------- Introduction -----------------
\paragraph*{Introduction.}
The decay of the three-dimensional (3D) Taylor--Green vortex (TGV) from a symmetric large-scale state into fully developed turbulence has long served as a canonical setting for probing the creation of small scales and the timing of peak dissipation \cite{TaylorGreen1937,Brachet1984,Pereira2021PRF}.
Spectral diagnostics, such as the energy spectrum $E(k)$, enstrophy spectrum $Z(k)\!\propto\!k^2E(k)$, and cumulative flux $\Pi(K)$, are the standard for characterizing interscale transfer, whereas real-space visualization often relies on the $Q$-criterion \cite{HuntWrayMoin1988,ChongPerryCantwell1990}.
However, reliable \emph{predictors} of when the dissipation surge will occur remain scarce; commonly tracked observables either evolve too slowly to be actionable (e.g., the development of inertial-range slopes in $E(k)$ \cite{DeBonis2013_TGV_NASA_TM,Brachet1984}) or exhibit finite propagation/lag across scales (e.g., flux-based indicators \cite{Cardesa2015_PoF_FluxTime,Valente2014_PRE_Imbalance}). Here, we report that the spectrum of $|\curl\boldsymbol{\omega}|^2$--equivalently, a $k^4$-weighted energy spectrum for incompressible flow ($\nabla\!\cdot\!\mathbf{u}=0$) because $\curl\boldsymbol{\omega}=-\nabla^2\mathbf{u}$--exhibits a sharply small-scale-biased and \emph{predictive} behavior.
Physically, the $k^4$ weight singles out the rapid growth of high-curvature vortex sheets and tightly wound tubes; as these structures form, the high-$k$ front of $|\curl\boldsymbol{\omega}|^2$ advances and its peak wavenumber then levels off \emph{before} viscous action is maximized.
Direct numerical simulations of the TGV up to $1024^3$ show that this precursor consistently predicts the time of the maximum $\varepsilon(t)$ and precedes the largest peak scale of $|\Pi|$, establishing the robust ordering that underpins our main result.

% ---------------- Added texts ------------------
%\paragraph*{Context and Reynolds-number perspective.}
\HL{In canonical high--Reynolds--number turbulence, the mean dissipation rate often follows
$\varepsilon \sim C_\varepsilon U^3/L$ with only a weak dependence of $C_\varepsilon$ on $Re$ in stationary,
equilibrium--cascade settings, whereas non--equilibrium transients can deviate from this classical picture
(e.g.\ \cite{Vassilicos2015,AlexakisBiferale2018}).
From a Reynolds-number viewpoint, it is then natural to seek diagnostics that are anchored in the physics of the
small scales. Our diagnostic is a simple $k^4$--weighting of the energy spectrum, i.e.\ $|\curl\boldsymbol{\omega}|^2(k)=k^4E(k)$
for incompressible flow (since $\curl\boldsymbol{\omega}=-\nabla^2 \mathbf{u}$), which privileges the incipient intensification of curvature
in sheets and tightly wound tubes. This curvature--biased sensitivity is consistent with cascade--centric viewpoints and with
scenarios in which vortex interaction and reconnection events accelerate small--scale formation \cite{YaoHussain2022}.
While we demonstrate the mechanism in the decaying Taylor--Green vortex, the construction itself is agnostic to the initial
condition and amenable to examination across other transients and $Re$.}

% ---------------- Figures -----------------
\begin{figure*}[t]
  \centering
  \includegraphics[width=\linewidth]{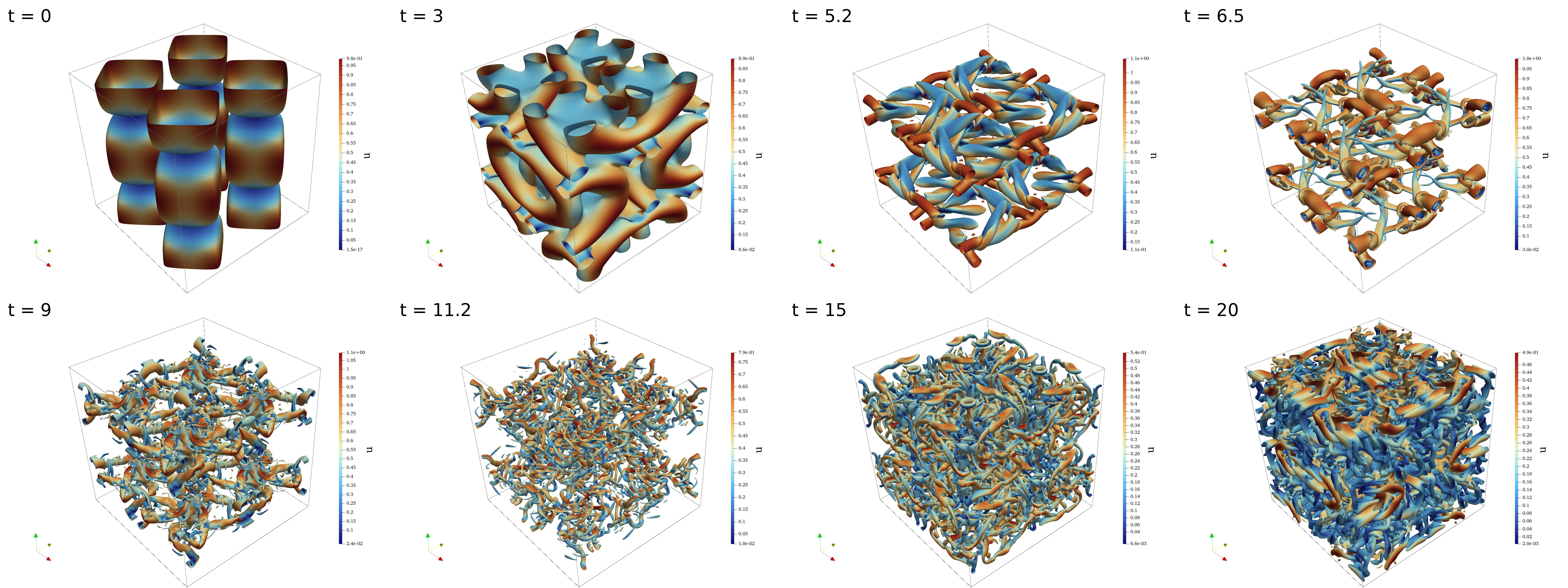}
  \caption{$Q$-criterion isosurfaces (colored by $|\bfu|$) at selected times
  for the $512^3$ run ($t=0,\,3,\,5.2,\,6.5,\,9,\,11.2,\,15,\,20$).
  The morphological transition to filament-dominated structures occurs prior to the dissipation peak,
  consistent with the spectral precursor identified in Fig.~\ref{fig:fig_timeval_kpeak}.}
  \label{fig:fig_snapshot}
\end{figure*}

% ---------------- Methods -----------------
\paragraph*{Methods.}
We solve the incompressible Navier--Stokes equations in a periodic cube $[0,2\pi]^3$ with viscosity $\nu=10^{-3}$ and the TGV initial condition $\bfu(x,y,z,0)=(\sin x\cos y\cos z,\,-\cos x\sin y\cos z,\,0)$.
A massively parallel pseudo-spectral solver with FFTW--MPI slabs~\cite{FFTW3} evaluates nonlinear terms in physical space, applies de-aliasing using a two-thirds \emph{box} cutoff together with the Helmholtz projection in Fourier space, and advances in time using a four-stage Runge--Kutta scheme and an integrating factor for viscosity~\cite{PattersonOrszag1971,CanutoBook}.
Simulations at $N^3=256^3,512^3,1024^3$ are performed with $\Delta t=10^{-3}$ up to $t=20$, writing velocity, and vorticity snapshots at regular intervals for post-processing.
The code uses the real-to-complex layout of FFTW with the required $N\!+\!2$ padding on the last dimension; snapshots are stored per rank to enable parallel I/O and exact reconstruction.

From the snapshots, we compute the isotropic shell averages of the energy spectrum $\E(k)$, enstrophy spectrum $\Z(k)$, dissipation density $\D(k)=2\nu k^2\E(k)$, nonlinear transfer $T(k)$, cumulative flux $\PiC(K)=-\sum_{m\le K}T(m)$, and spectrum of $|\curl\bfw|^2$ (for incompressible flow equivalent to $k^4\E(k)$).
Shells are defined by integer bins of $k=\mathrm{round}(|\boldsymbol{k}|)$ and plane weights $w_z=1$ are used for $k_z\in\{0,N/2\}$ and $w_z=2$; otherwise, the FFT normalization is $1/N^3$ such that $\sum_k \E(k)$ is equal to the volume-averaged kinetic energy.
For the analysis, we apply a \emph{spherical} two-thirds mask and record it as a binary column ``mask'' in the CSV; this differs from the \emph{box} cutoff enforced during time marching. 
Conservation diagnostics verify $\sum_{k}T(k)\approx0$ and $\Pi(K_{\max})\approx0$; specifically, we confirm $\bigl|\sum_{k}T(k)\bigr|\le 1.5\times10^{-8}$ and $\bigl|\Pi(K_{\max})\bigr|\le 1.0\times10^{-7}$. 
Using a spherical two-thirds analysis mask, the maximum relative imbalance at the dissipation peak $(t=t_\varepsilon)$ is 
$\bigl|\sum_{k}T\bigr| \big/ \sum_{k}\!\bigl|T\bigr|\ \le 10^{-16}$ and 
$\bigl|\Pi(K_{\max})\bigr| \big/ \max_{K}\!\bigl|\Pi\bigr|\ \le 10^{-16}$. 
In addition, we monitor $\langle \bfu\cdot(\bfu\cdot\nabla)\bfu\rangle$ and the RMS divergence $\|\nabla\!\cdot\!\bfu\|_{\mathrm{rms}}$ in physical space to ensure numerical consistency~\cite{CanutoBook}.
The Q-criterion isosurfaces (colored by $|\bfu|$) are produced using a scripted ParaView pipeline~\cite{Ahrens2005ParaViewAE}.

\paragraph*{Transfer function $T(k)$.}
The shell energy budget can be expressed as follows:
\begin{equation}
\partial_t E(k,t) \;=\; T(k,t) - 2\nu k^2 E(k,t).
\label{eq:shellbudget}
\end{equation}
However, in practice, we compute $T(k,t)$ \emph{directly} from the modal transfers in Fourier space:
\begin{equation}
T(k,t)
  = - \sum_{\boldsymbol{\kappa}\in\mathcal{S}_k}
      {\mathfrak R}\!\left\{
        \hat{\bfu}^{\ast}(\boldsymbol{\kappa},t)\cdot
        \widehat{\mathbf{N}}_{\perp}(\boldsymbol{\kappa},t),
      \right\},
\label{eq:Tkdef}
\end{equation}
where the summation runs over all Fourier modes:
$\boldsymbol{\kappa}\in\mathcal{S}_k=\{\boldsymbol{\kappa}\,:\,|\boldsymbol{\kappa}|=k\}$,
i.e.,\ all wavevectors of magnitude~$k$~\cite{WaleffeFabian1992}.
Here, $\mathbf{N}=(\bfu\!\cdot\!\nabla)\bfu$ and
$\widehat{\mathbf{N}}_{\perp}(\boldsymbol{\kappa})=
\mathbf{P}(\boldsymbol{\kappa})\,\widehat{\mathbf{N}}(\boldsymbol{\kappa})$
uses the Leray projector
$\mathbf{P}(\boldsymbol{\kappa})=
\mathbf{I}-\boldsymbol{\kappa}\boldsymbol{\kappa}^\top/|\boldsymbol{\kappa}|^2$
to enforce incompressibility and remove pressure~\cite{DomaradzkiRogallo1990,AlexakisBiferale2018}.
In the pseudo-spectral implementation, $\mathbf{N}$ is assembled in physical space from $u_i$ and the spectral derivatives
($\partial_j u_i \leftrightarrow i\kappa_j\hat u_i$), and then transformed into Fourier space; the optional spherical two-thirds analysis cutoff and the projection $\mathbf{P}$
are applied before forming \eqref{eq:Tkdef}.
The isotropic shells $\mathcal{S}_k$ are defined by rounding $|\boldsymbol{\kappa}|$ to the nearest integer; the $k_z=0$ and $k_z=N/2$ planes carry a unit weight, and the others carry a weight $2$.
We accumulate the shell sums using the Kahan summation~\cite{NJHigham2002KahanSum} and report the cumulative flux $\Pi(K)=-\sum_{m\le K}T(m)$.
With these conventions, the discrete conservation properties hold to numerical precision, i.e., $\sum_k T(k)\approx 0$ and $\Pi(K_{\max})\approx 0$, as noted above.

\paragraph*{Scale measures and physical meaning.}
We monitor four complementary scale measures together with the dissipation rate.
First, the dissipation rate
\[
\varepsilon(t) \;=\; \sum_{k} \D(k,t) \;=\; 2\nu \sum_{k} k^2 \E(k,t)
\]
is the mean kinetic energy loss per unit time, which is equivalent to $2\nu\langle S_{ij}S_{ij}\rangle$ in physical space, and sets the instantaneous Kolmogorov scale
\[
\eta(t) \;=\; \Big(\nu^3/\varepsilon(t)\Big)^{1/4}, \qquad k_\eta(t)\equiv \eta(t)^{-1},
\]
which characterizes the viscous cutoff of the cascade.
Second, the \emph{integral scale}
\[
L_{\mathrm{int}}(t) \;=\; \frac{\pi}{\,2\sum_k \E(k,t)\,}\,
\sum_{k\ge 1} \frac{\E(k,t)}{k}
\]
weights the spectrum towards the energy-containing range and tracks the progressive erosion of large-scale coherence as the energy spreads to a higher $k$.
Third, the \emph{Taylor microscale}
\[
\lambda(t) \;=\; \left(\frac{\sum_k \E(k,t)}{\sum_k k^2 \E(k,t)}\right)^{1/2}
\]
captures an intermediate characteristic length between $L_{\mathrm{int}}$ and $\eta$, and is the basis for the Taylor-scale Reynolds number $R_\lambda$~\cite{Ishihara2009}.
Finally, we define the curvature-biased indicator
\begin{align}
k_{\mathrm{peak}}\!\left[\,|\curl\bfw|^2\,\right](t)
\;&\equiv\;
\arg\max_{k\ge 1}\big\{\,|\curl\bfw|^2(k,t)\,\big\}, \nonumber \\
|\curl\bfw|^2(k,t)\;&=\;(k^4)\,\E(k,t) \nonumber
\end{align}
for incompressible flow (because $\nabla \times \bfw
= \nabla \times (\nabla \times \bfu)
= \nabla(\nabla \cdot \bfu) - \nabla^{2} \bfu
= -\nabla^{2} \bfu$ when $\nabla\cdot\bfu=0$).
Owing to its $k^4$ weighting, $|\curl\bfw|^2(k)$ responds earliest to the formation of high-curvature sheets and tightly wound tubes, rapid advance, and subsequent leveling off of $k_{\mathrm{peak}}[\,|\curl\bfw|^2\,]$, thus providing a small-scale-biased precursor to the dissipation surge.
All spectral sums are isotropic shell averages with the usual plane weights for real-to-complex FFTs ($k_z=0$ and $k_z=N/2$ are weighted by $1$ and others by $2$); when shown, the analysis plots are restricted to shells with \texttt{mask}$=1$ (a spherical two-thirds cutoff).

\begin{figure}[t]
  \centering
  \includegraphics[width=0.98\linewidth]{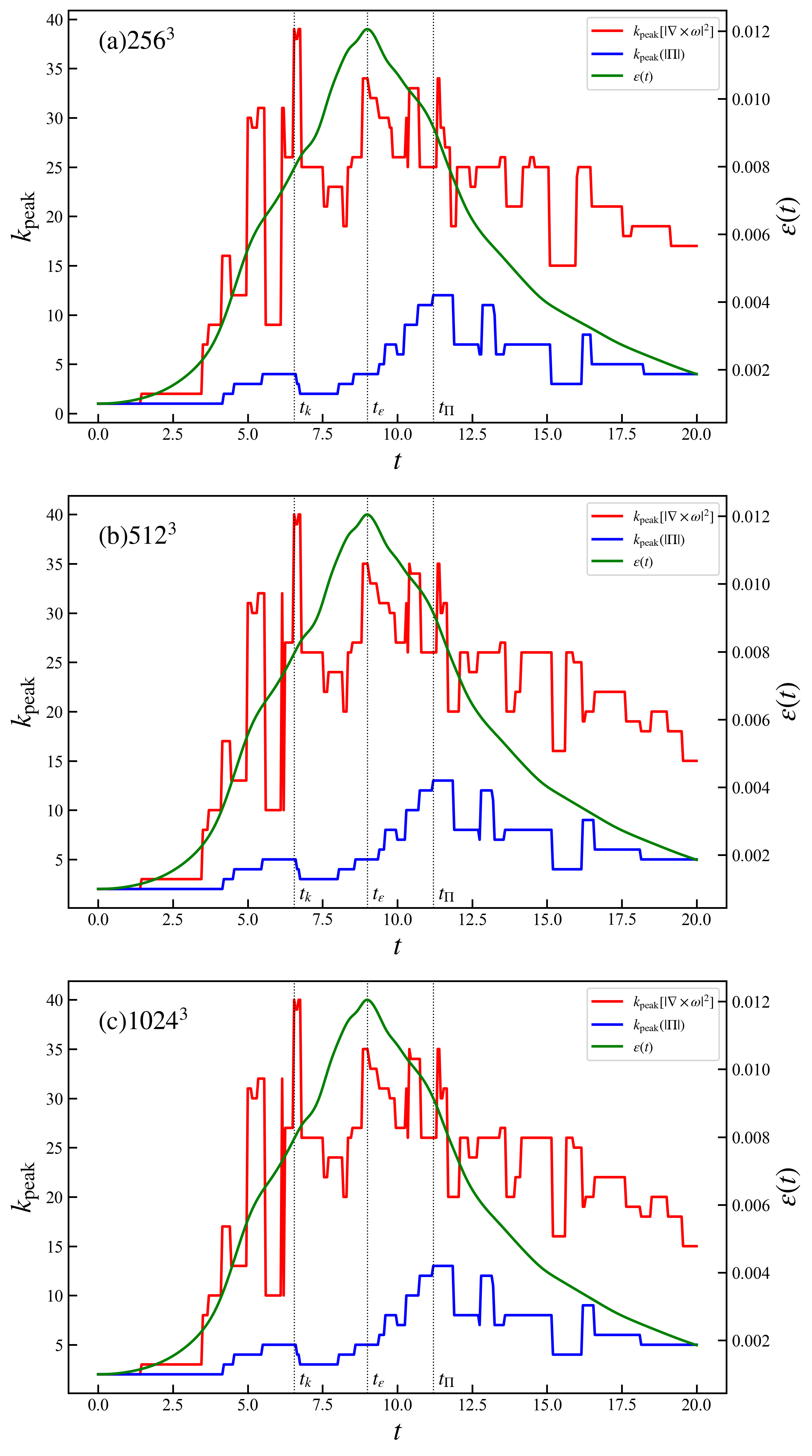}
  \caption{\textbf{Main result.} Time evolution of (red) $k_{\mathrm{peak}}\big[\,|\curl\bfw|^2(k)\,\big]$, (blue) $k_{\mathrm{peak}}\big(|\PiC(K)|\big)$, and (green) dissipation $\varepsilon(t)=\sum_k 2\nu k^2 \E(k)$ for $N^3=256^3,512^3,1024^3$.
  Dashed verticals indicate $t_k$ (red), $t_\varepsilon$ (green), and $t_\Pi$ (blue).
  The consistent ordering $t_k < t_\varepsilon < t_\Pi$ is evident and almost grid independent.}
  \label{fig:fig_timeval_kpeak}
\end{figure}

\begin{figure}[t]
  \centering
  \includegraphics[width=0.95\linewidth]{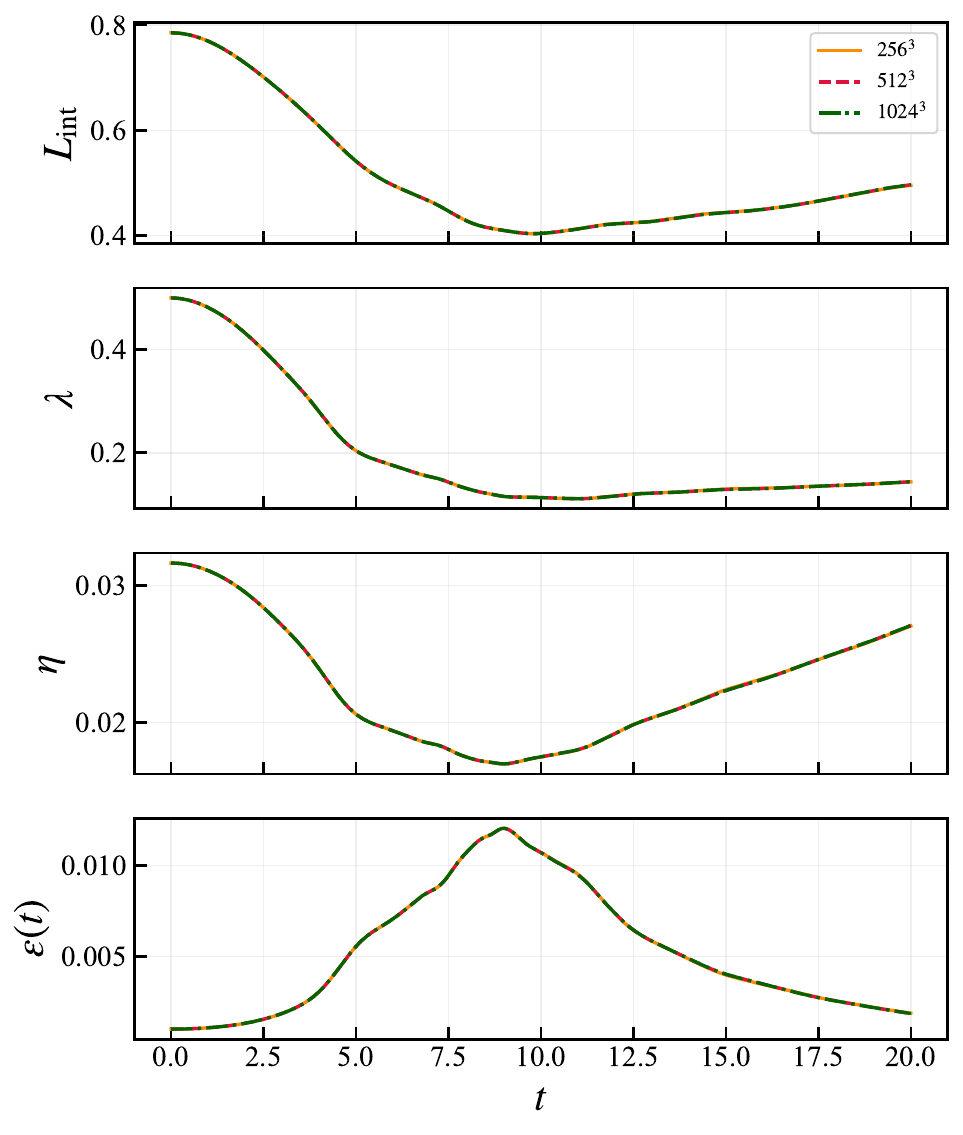}
  \caption{Integral scales: $L_{\mathrm{int}}(t)=\frac{\pi}{2\sum_k \E(k)}\sum_k \E(k)/k$, $\lambda(t)=\sqrt{\sum_k \E(k)/\sum_k k^2 \E(k)}$, $\eta(t)=(\nu^3/\varepsilon)^{1/4}$, and $\varepsilon(t)$. In Fig.~\ref{fig:fig_timeval_kpeak}, the peak wavenumber $k_{\mathrm{peak}}[\,|\curl\bfw|^2\,]$ increases rapidly and levels off by $t\simeq 6.5$, preceding the minimum of $\eta$ (i.e.,\ the dissipation peak).}
  \label{fig:fig_timeval_scales}
\end{figure}

\begin{figure*}[t]
  \centering
  \includegraphics[width=\linewidth]{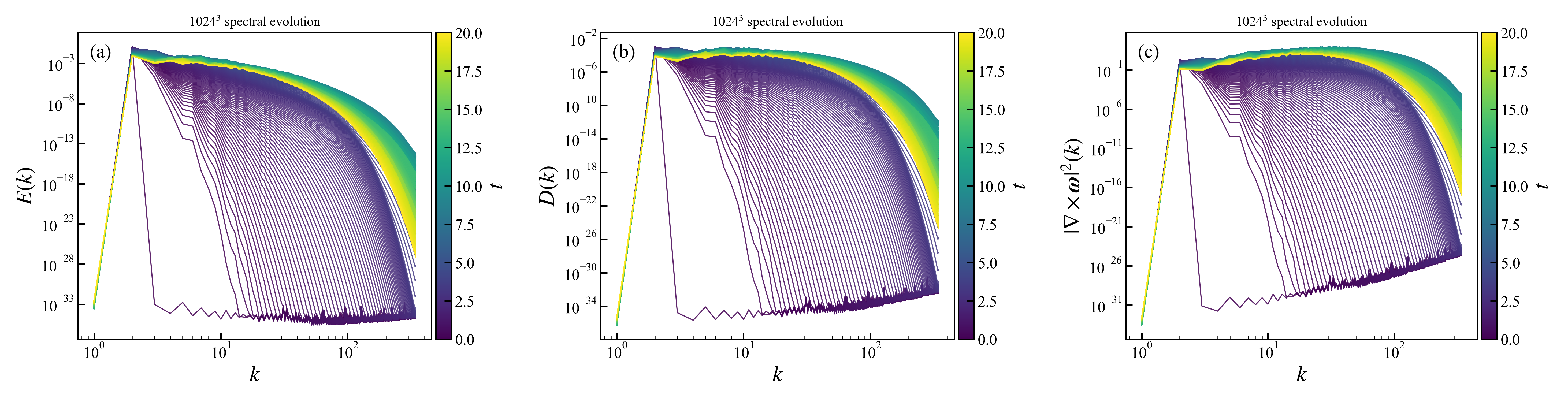}
  \caption{\textbf{Spectral evolution at $1024^3$.}
  (a) $E(k)$, (b) $D(k)=2\nu k^2E(k)$, and (c) $|\nabla\times\boldsymbol{\omega}|^2(k)$
  plotted for all saved times $t\in[0,20]$ with color encoding $t$.
  Only shells with \texttt{mask} = 1 (spherical two-thirds analysis cut) are shown. The high-$k$ front of $|\curl\boldsymbol{\omega}|^2$ advances first and then levels off earlier than that of $D(k)$, thereby anticipating the dissipation peak (cf.\ Fig.~\ref{fig:fig_timeval_kpeak}).}
  \label{fig:fig_spectra_evolution}
\end{figure*}

\begin{figure}[t]
  \centering
  \includegraphics[width=\columnwidth]{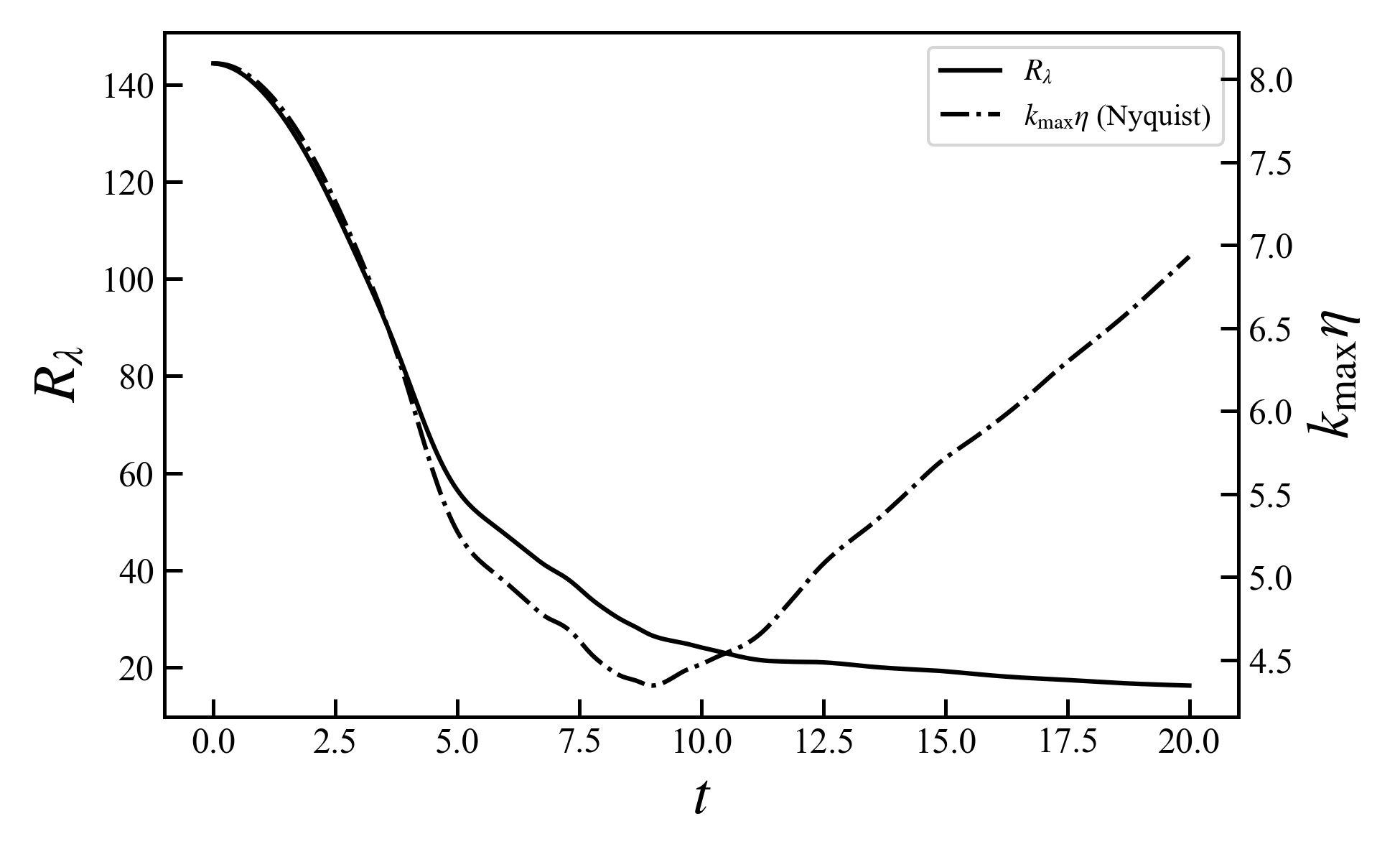}
  \caption{Resolution metrics for $512^3$: time series of $R_\lambda(t)$ and
  $k_{\max}\eta(t)$ (Nyquist definition). The spectra are calibrated to $K_0=1/8$ and $\varepsilon(t)$ is computed from
  $D(k)=2\nu k^2E(k)$.}
  \label{fig:figA1_Rlambda_kmaxeta_512}
\end{figure}

\begin{figure}[t]
  \centering
  \includegraphics[width=\columnwidth]{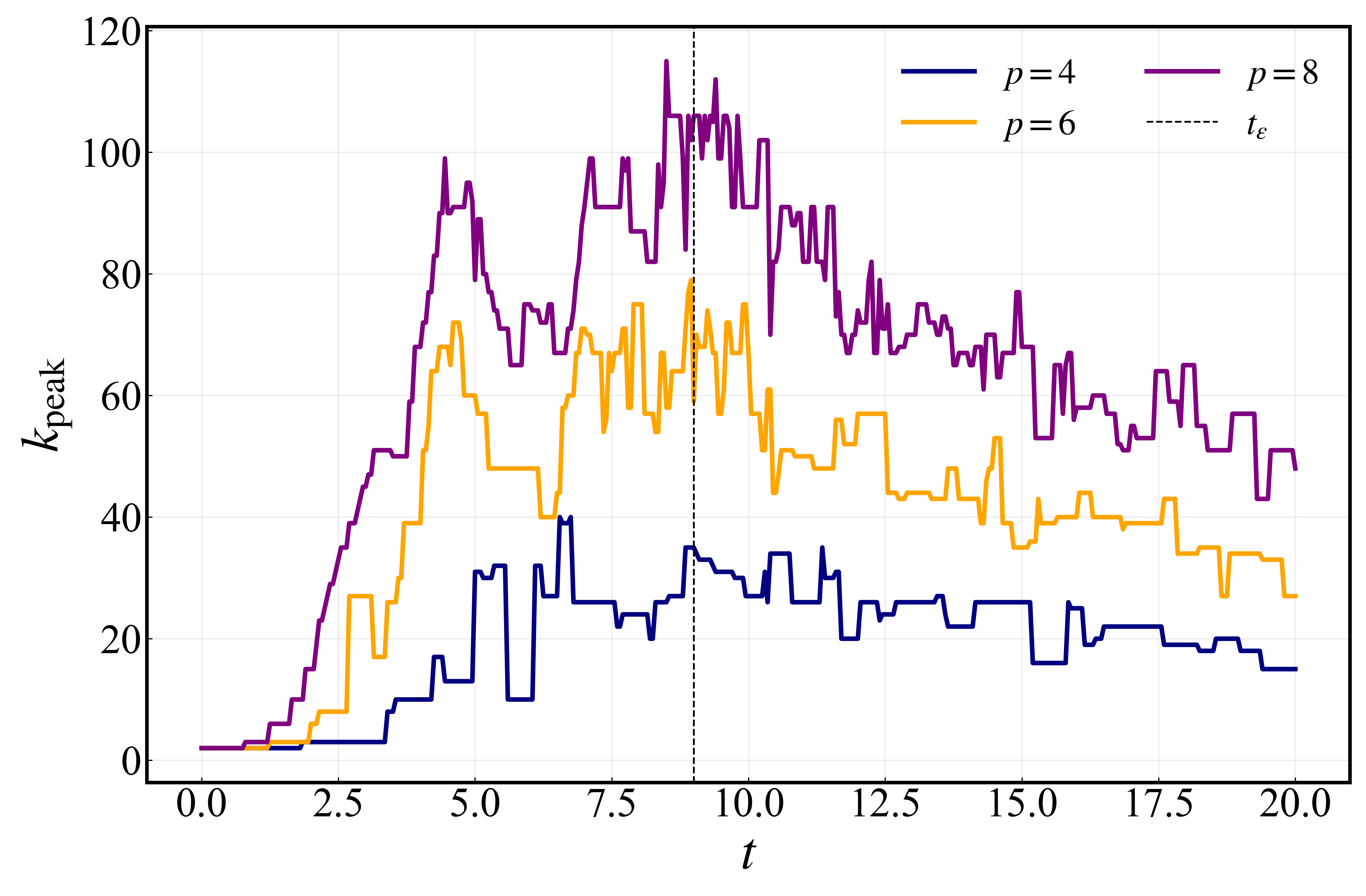}
\caption{\HL{Time evolution of $k_{\mathrm{peak}}[k^{p}E(k)]$ for $p=4,6,8$ in the $1024^{3}$ run. Both $p=6$ and $p=8$ peaks appear after the $p=4$ peak and before $t_{\varepsilon}$; the brief pre-$t_{\varepsilon}$ spike for $p=6$ is transient, confirming that increasing $p$ does not yield a monotonic advance and supporting $p=4$ as the most robust indicator.}}
  \label{fig:figA2_kpeaks_1024_higherorders}
\end{figure}

\paragraph*{Results.}
Figure~\ref{fig:fig_snapshot} (for $512^3$) shows the $Q$-isosurfaces at representative times, illustrating the transition from large-scale structures to a filament-dominated state prior to the dissipation peak.
For all resolutions, we verify that the runs are well resolved at the time of peak dissipation.
Using the Nyquist definition $k_{\max}=(N/2)\,2\pi/L$, the resolution metric $k_{\max}\eta(t)$ remains above unity throughout and reaches $\simeq 4.35$ at $t_\varepsilon$ for the $512^3$ case, with comparable or larger values at $1024^3$ (see Appendix, Fig.~\ref{fig:figA1_Rlambda_kmaxeta_512}).
The Taylor-scale Reynolds number $R_{\lambda}(t)$ decreases as dissipation grows and small scales form, consistent with the evolution of $\varepsilon(t)$ (Appendix Fig.~\ref{fig:figA1_Rlambda_kmaxeta_512}).

The main results are summarized in Fig.~\ref{fig:fig_timeval_kpeak}.
The peak wavenumber of $|\curl\bfw|^2$ (red) surges to intermediate-small scales with an onset at time $t_k$, and then levels off.
The dissipation rate $\varepsilon(t)$ (green) reaches its maximum a time $\Delta t\!\approx\!2$--$3$ later.
The peak scale of the flux (blue) typically occurs even later ($t_\Pi$), establishing a robust ordering $t_k<t_\varepsilon<t_\Pi$ across $256^3$, $512^3$, and $1024^3$.
This ordering relates the onset of curvature intensification (Fig.~\ref{fig:fig_snapshot}) to the subsequent dissipation surge and, only thereafter, to the largest-flux scale.
The integral scales in Fig.~\ref{fig:fig_timeval_scales} show that $L_{\mathrm{int}}$ and $\lambda$ decrease as $t\!\sim\!6.5$, whereas the Kolmogorov scale $\eta=(\nu^3/\varepsilon)^{1/4}$ reaches its minimum near $t_\varepsilon$~\cite{Kolmogorov1991,Frisch1995}.
Therefore, the early stabilization of $k_{\mathrm{peak}}[\,|\curl\bfw|^2\,]$ signals the imminent action of viscosity on the newly formed high-curvature structures.

To understand \emph{why} the precursor in Fig.~\ref{fig:fig_timeval_kpeak} emerges, we inspect the full time histories of the spectra at $1024^3$ (Fig.~\ref{fig:fig_spectra_evolution}).
Panel (a) shows $E(k)$: energy spreads steadily from the initial large scales towards higher $k$, with the spectral front advancing up to the analysis mask near the dissipation peak.
Panel (b) shows $D(k)=2\nu k^2E(k)$, a $k^2$-weighted version of $E(k)$ that emphasizes the intermediate wavenumbers earlier than $E(k)$ itself.
Most strikingly, panel (c) shows $|\curl\boldsymbol{\omega}|^2(k)$, which is effectively a $k^4$-weighted energy spectrum in incompressible flow, whose \emph{high-$k$ front} surges first and then levels off by $t\simeq t_k$ (cf.\ Fig.~\ref{fig:fig_timeval_kpeak}).
After this early leveling off, the spectrum increases primarily in amplitude rather than in extent, and the dissipation rate reaches its maximum only later, at $t_\varepsilon$, followed by the largest-flux peak scale at $t_\Pi$.
Therefore, the ordering $t_k<t_\varepsilon<t_\Pi$ is a direct consequence of the progressively stronger small-scale weights ($k^0\!\to\!k^2\!\to\!k^4$) that select, in turn, curvature intensification, viscous action, and net interscale transport.
All curves use standard shell averaging; only shells with \texttt{mask}$=1$ (a spherical two-thirds analysis cut) are shown (see Methods).

\paragraph*{\HL{Comparison across $k^{p}E(k)$.}}
\HL{We further examined the family $k^{p}E(k)$ for $p=6,8$
(Appendix Fig.~\ref{fig:figA2_kpeaks_1024_higherorders}).
For both $p=6$ and $p=8$, the appearance of their peaks precedes the dissipation maximum $t_\varepsilon$
but occurs later than the $p=4$ peak.
These peaks are brief and less well defined, and it is difficult to discern whether their sharpness
reflects genuine dynamics or residual spectral noise near the cut-off.
Thus, increasing $p$ does not yield a monotonic advance of the characteristic peak time;
rather, higher-order weights amplify finite-resolution sensitivity.
Among the tested exponents, $p=4$ remains the most robust and interpretable choice,
possibly because it accentuates the formation of high-curvature structures
(\,$|\curl\boldsymbol{\omega}|^2 \equiv k^4E(k)$ for incompressible flow\,)
that mark the onset of strong viscous action.
Hence, $p=4$ represents a practical ``sweet spot'' between early responsiveness and robustness for DNS post-processing.}

\paragraph*{\HL{Scope and outlook.}} 
\HL{The claims of this Letter are established for the decaying TGV at $\nu=10^{-3}$. We focus on the delay $\Delta t$ between the stabilization of $k_{\mathrm{peak}}\!\left[\,|\curl\boldsymbol{\omega}|^2\,\right]$ and the dissipation--peak time $t_\varepsilon$. By dimensional reasoning, $\Delta t$ is expected to be $O(1)$ when normalized by the Kolmogorov time $\tau_\eta=(\nu/\varepsilon)^{1/2}$, and thus only weakly dependent on $Re$, whereas normalization by a large--eddy time $T_L$ may reflect non--equilibrium features of the transient. Systematic tests across different initial conditions (e.g.\ random--phase low--$k$, Kida--Pelz, ABC) and across $Re$ as well as forced/maintained turbulence will be valuable next steps; practically, the diagnostic lends itself to adaptive meshing and output scheduling because it is computed at negligible cost from standard spectra.}

\paragraph*{Conclusion.}
The peak wavenumber of the $|\curl\bfw|^2$ spectrum provides a simple and robust early-warning signal for the dissipation surge in a 3D TGV.
Because it is simply a higher-order ($k^4$) weighting of $\E(k)$, the diagnostic is inexpensive to compute, yet strongly emphasizes the regions in which the curvature intensifies and viscous effects will soon dominate.
We anticipate applications to adaptive meshing and scheduling, as well as to comparative studies of non-equilibrium transients in different initial conditions or forcing protocols.

\paragraph*{Appendix: Resolution metrics.}
We monitor the resolution using the Nyquist definition $k_{\rm max} =(N/2)2\pi/L$ and report the time series of $k_{\rm max}\cdot\eta(t)$ together with the Taylor-scale Reynolds number $R_{\lambda}(t) ={u}^{\prime}(t)\lambda(t)/\nu$. Here, $\eta=(\nu^{3}/\varepsilon)^{1/4}$, ${u^{\prime 2}}=\frac{2}{3}\sum_{k}E(k)$, and $\lambda=\sqrt{\sum_k E(k)/\sum_k k^{2}E(k)}$, with $k$ the physical wavenumber $(2\pi/L)\times$ the integer shell index. All spectra are calibrated such that the total energy at the first snapshot is equal to $K_{0}=1/8$ (the Taylor--Green initial condition), and $\varepsilon(t)$ is obtained from $D(k)=2\nu k^2E(k)$. The shell sums use the spherical two-thirds mask, unless stated otherwise. Across all resolutions, $k_{\rm max}\cdot \eta(t)$ remains comfortably above unity at all times and attains $k_{\rm max}\eta(t)\simeq 4.35$ at the dissipation peak for $512^3$; therefore, the $1024^3$ run meets the stringent resolution criterion. The concomitant $R_{\lambda}(t)$ shows a decrease as dissipation intensifies and small scales develop, consistent with the evolution of $\varepsilon(t)$.

\paragraph*{Acknowledgments.}
This study was supported by JSPS KAKENHI (Grant Number 22K14177) and JST PRESTO (Grant Number JPMJPR23O7).

% ---------------- References -----------------
\bibliographystyle{apsrev4-2}
\bibliography{references}

\end{document}